# Synthesis and structure of new pyrochlore-type oxides $Ln_2ScNbO_7$ (Ln = Pr, Nd, Eu, Gd, Dy)

S. Zouari[a,b], R. Ballou [b], A. Cheikh-Rouhou [a], P. Strobel [b]

[a]Laboratoire de Physique des Matériaux, Faculté des Sciences de Sfax, BP 763, 3038 Sfax, Tunisia

[b]Institut Néel, CNRS, B. P. 166, 38042 Grenoble Cedex 9, France

**Abstract**

We report the synthesis and structural study of mixed oxides in the $Ln_2ScNbO_7$ series. New phases with Ln = Pr, Eu, Gd and Dy are obtained. All crystallize in the cubic pyrochlore structure type, space group F-d3m, with no Sc-Nb ordering on the B-site. The structures are determined by Rietveld refinement. The evolution of cell parameters, interatomic distances and angles as a function of lanthanide cation size is discussed. Magnetic measurements show the absence of ordering down to 2 K, in agreement with the presence of strong geometric frustration in the lanthanide sublattice. The europium phase shows a peculiar magnetic behaviour; its magnetic susceptibility becomes constant below ca 50 K. This feature confirms the behaviour observed previously on $Eu_2Ti_2O_7$ and is ascribed to crystal field effects.

Corresponding author: P. STROBEL

Institut Néel, CNRS

Case F, BP 166, 38042 Grenoble Cedex 9, France

Phone: 33+ 476 887 940

FAX: 33+ 476 881 038

E-mail:  pierre.strobel@grenoble.cnrs.fr

**Introduction**

Pyrochlore-type oxides $A_2B_2O_7$ attracted much interest recently because they present a high geometric frustration on both A and B sublattices [1,2]. In this structure, the A site is usually occupied by large cations such as lanthanides (Ln), whereas the B site fits better smaller first- or second-row transition elements. The most favourable situation for studies of lanthanide magnetism arises when the Ln cation is combined with a diamagnetic $B^{4+}$ cation. In an $A^{3+}_2B^{4+}_2O_7$ pyrochlore formula, the choice is then limited to $Ti^{4+}$ and $Sn^{4+}$ (and marginally $Zr^{4+}$ or $Ge^{4+}$, which form pyrochlore compounds with fewer lanthanides). $Ln_2Ti_2O_7$ pyrochlores have been widely studied [2,3].

It is always interesting to find new examples of compounds containing lanthanides in a highly frustrated configuration, especially for early rare earths, since the titanium pyrochlores do not form with Ln = Pr or Nd. We consider here pyrochlore compositions where $B^{4+}_2$ is replaced by a $B^{3+}B'^{5+}$ combination where B and B' have suitable sizes and carry no spin. Few examples of such formulas have been reported so far : in the Sc-Nb system, only the neodymium [4] and samarium [5] compounds were described previously, and they were characterized by their cell parameters only. In this paper, we describe the preparation and structural properties of an extended series of $Ln_2ScNbO_7$ pyrochlore compositions with Ln = Pr-Dy. Preliminary magnetic measurements are briefly presented; a more comprehensive description of their magnetic properties will be given elsewhere [6].

**Experimental**

Powder samples of rare earth pyrochlore oxides were prepared by solid state reaction, according to the following reactions :

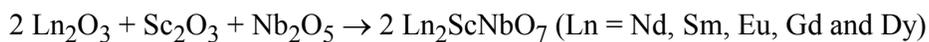
$2\ Ln_2O_3 + Sc_2O_3 + Nb_2O_5 \rightarrow 2\ Ln_2ScNbO_7$ (Ln = Nd, Sm, Eu, Gd and Dy)

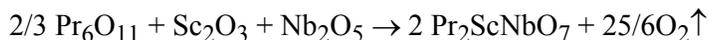
$2/3\ Pr_6O_{11} + Sc_2O_3 + Nb_2O_5 \rightarrow 2\ Pr_2ScNbO_7 + 25/6 O_2\uparrow$

Starting reagents were 99.9 % binary oxides. The rare earth oxides were dried before use. Stoichiometric mixtures were intimately mixed in an agate mortar and heated repeatedly at 1400 °C in air, with intermittent regrinding and pelletizing.

Phase purity, homogeneity and unit cell dimensions were determined by powder X-ray diffraction (XRD) in transmission geometry, using a Bruker D8 diffractometer equipped with an incident-beam monochromator. X-ray patterns were recorded with Cu-K$a$ radiation in the range 26–96° using a step size of 0.02 ° and counting rate 80 s per step. Structures were refined using room-temperature X-ray data by the Rietveld method, using the *Fullprof* program [7].

Magnetization measurements were carried out in an extraction magnetometer in the temperature range 2–300 K, using magnetic field up to 10 T.

**Results and discussion**

All $Ln_2ScNbO_7$ compositions but one gave X-ray patterns indexable in a face-centered, pyrochlore-type cubic structure. The exception is Sm, which was not obtained as single-phase material. Figure 1 shows a portion of the X-ray powder diffraction patterns of selected $Ln_2ScNbO_7$ compounds, emphasizing the difference between phase-pure Nd and Eu phases and $Sm_2ScNbO_7$, which contains significant amounts of $SmNbO_4$ [8] in spite of similar preparation conditions.

For the sake of completion, it can be added that the synthesis of the member of this series with Ln = La failed. This is consistent with previous studies, which showed that $La_2B_2O_7$ pyrochlores form only with rather large $B^{4+}$ cations (Sn, Zr, Pb), but not with Mn, Ti, Ru or Mo [3].

The structures of new $Ln_2ScNbO_7$ compounds were refined from powder XRD data in a pyrochlore structure model, i.e. space group Fd3m, Ln atoms on 16d site (½ ½ ½), Sc and Nb randomly distributed on 16c site (0 0 0), O1 on 48f site (x 1/8 1/8), and O2 on 8b site (3/8 3/8 3/8). The refinement included the following variables: scale factor, background polynomial coefficients, cell parameter a, O1 variable position x, isotropic atomic displacement parameters, and the usual pseudo-Voigt profile parameters. No extra reflections attributable to a possible ordering of B and B' cations were observed.

The results are summarized in Table 1, together with selected interatomic distances and angles. An typical refinement is illustrated in Fig.2.

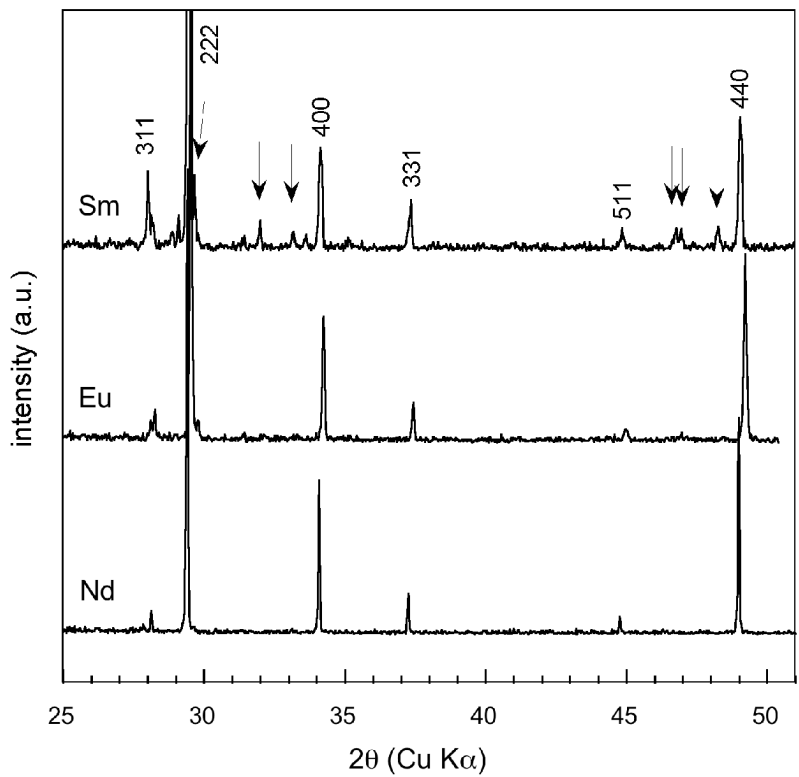

Fig.1. Showing a portion of XRD patterns of $Ln_2ScNbO_7$ phases with Ln = Nd, Sm and Eu. Arrows indicate the main reflections of $SmNbO_4$ impurity.

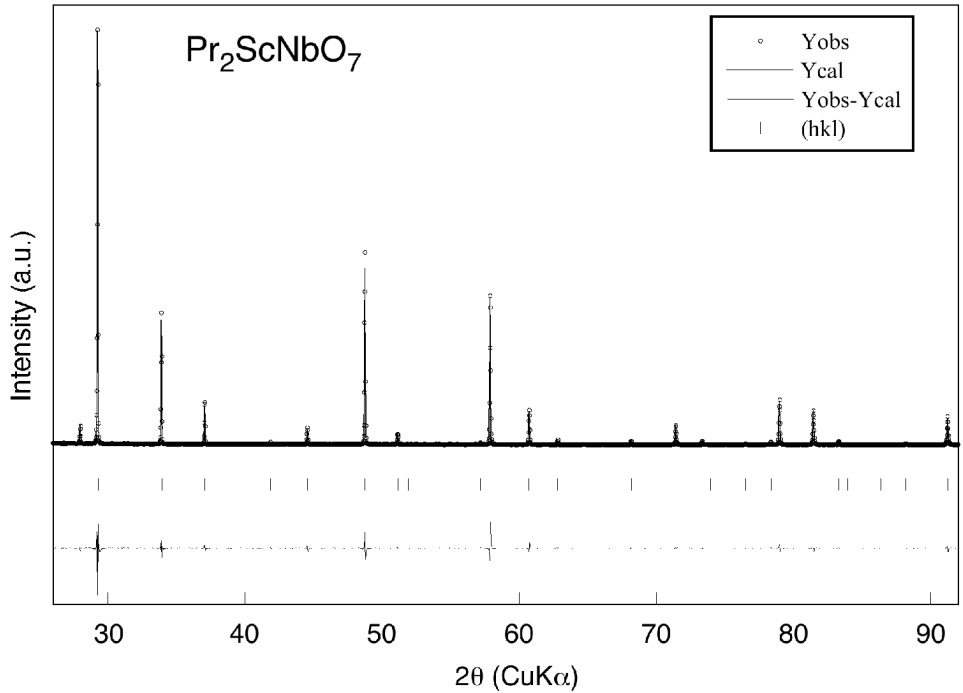

Fig.2. Observed (points) and calculated (continuous line) XRD pattern of $Pr_2ScNbO_7$. Pyrochlore reflexions are indicated by vertical bars. The bottom line is $I_{obs}-I_{calc}$.

The evolution of cell parameter $a$ and x(O1) (the only atomic coordinate variable) throughout the $Ln_2ScNbO_7$ series is shown in Fig. 3 as a function of Ln ionic radius $r_i$. As expected, $a$ increases with increasing $r_i$. A similar trend is observed on both Ln–O1 and Ln–O2 distances. Note that the B atom is coordinated only by O1, and that the B–O1 distance is almost constant throughout this series the series (see Table 1). This is consistent with the influence of the O1 atomic position, which varies in the direction opposite to that of the cell parameter (see Fig.3). It has been shown that an decrease in x(O1) induces a decrease in the B–O distance [9], and thus compensates for the cell volume increase with lanthanide cationic radius. The bonding angles (see Table 2) undergo only slight changes; they remain heavily distorted from a perfect cube around Ln atoms throughout the series.

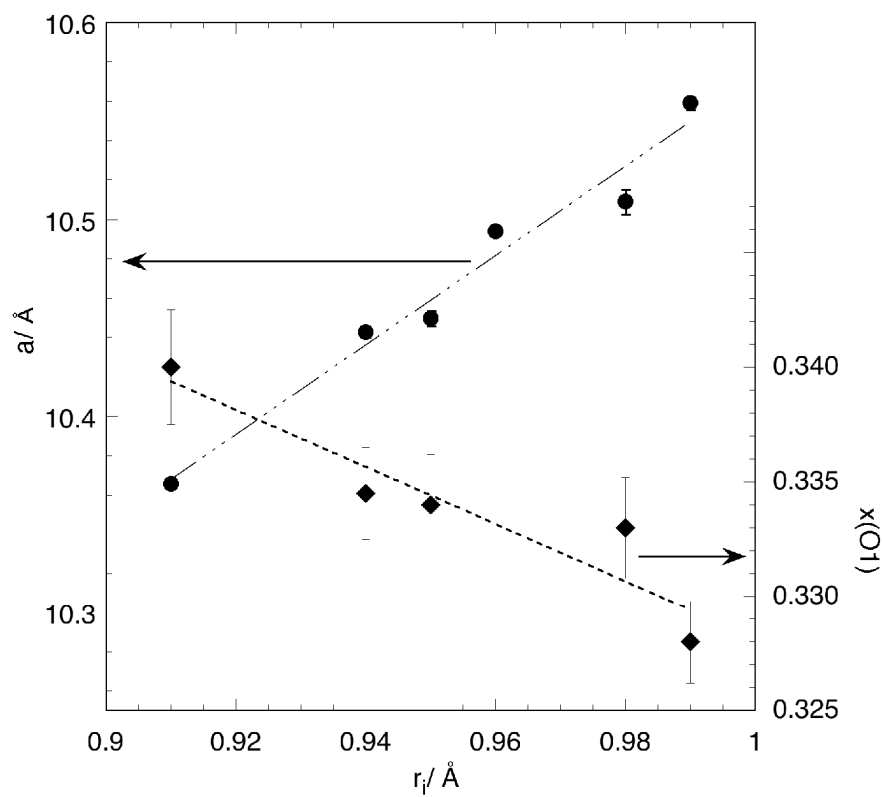

Fig.3. Evolution of cell parameter *a* (circles) and of O1 positional parameter x(O1) (diamonds) as a function of lanthanide ionic radius $r_i$ [10] throughout the $Ln_2ScNbO_7$ series.

Finally, note that the cell parameters throughout this series scale fairly well with those of $Ln_2B^{4+}{}_2O_7$ series [3] : they are closest to those of $Ln_2Sn_2O_7$, in agreement with the proximity of ionic radii of $Sn^{4+}$ (0.69 Å), $Nb^{5+}$ (0.68 Å) and $Sc^{3+}$ (0.74 Å) [10].

The temperature dependence of the magnetic susceptibility is shown for Ln = Nd and Eu in Fig.4. The neodymium case is typical of the series : no magnetic order sets in down to 2 K, although an extrapolation from the high-temperature region shows that the Weiss constant Q is very negative, showing dominant antiferromagnetic ordering. This behaviour is consistent with a heavily frustrated system. Measurements to lower temperatures may be necessary, since some pyrochlore-type rare-earth titanates or stannates have been found to order at temperatures ≤ 1.3 K [11-13]. The curvature of $\chi^-$

$^{1}$(T) at low temperature may have the same origin as that observed previously in titanates such as $(Y_{2-x}Tb_x)Ti_2O_7$, where it could be fitted assuming a doublet structure resulting from crystal field splitting [14].

The behaviour of $Eu_2ScNbO_7$ is markedly different : its magnetic susceptibility levels off on cooling to a temperature-independent value for T < ca. 40 K (Fig. 4). This feature is similar to that of the corresponding titanate $Eu_2Ti_2O_7$, where the constant susceptibility was ascribed to crystal field effects [15, 16]. The magnetic properties of the new pyrochlores obtained in this study will be analyzed in more detail elsewhere [6].

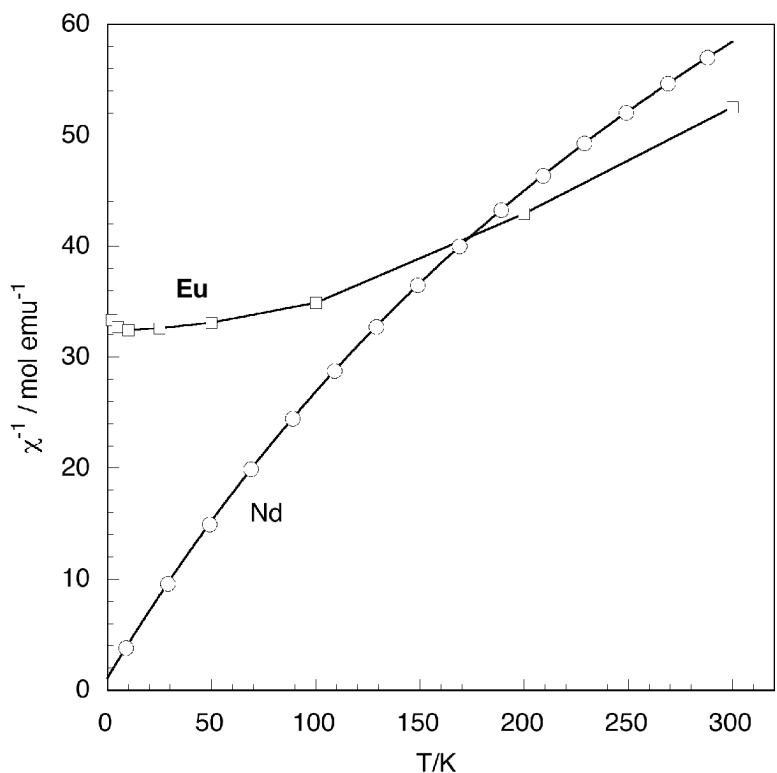

Fig.4. Temperature dependence of the reciprocal magnetic susceptibility $1/\chi$ for $Ln_2ScNbO_7$ phases with Ln = Nd (circles) and Eu (squares).

**Conclusions**

In this study, we were able to extend substantially the range of known rare-earth pyrochlore-type oxides by replacing the tetravalent B cation by a Sc-Nb combination. Four new compositions with Ln = Pr, Eu, Gd and Dy are reported, and single-phase $Ln_2ScNbO_7$ pyrochlores are now known to form from Ln = Pr to Dy. The structural properties are consistent with steric considerations based on the influence of the lanthanide ionic radius, while the O1 position is slightly modified to maintain the (Sc,Nb)-O fairly constant. No magnetic ordering is observed down to 2 K, in agreement with a highly frustrated configuration of the lanthanide sublattice.

**Acknowledgments**

We wish to thank Région Rhône-Alpes for financial support of S. Zouari and A. CHeikh-Rouhou (MIRA program).

Table 1. Structural parameters for single-phase $Ln_2ScNbO_7$ compounds from Rietveld refinements of powder X-ray data.

| Ln | Pr | Nd | Eu | Gd | Dy |
|---|---|---|---|---|---|
| $a_{cub}$ (Å) | 10.5591(2) | 10.5091(2) † | 10.4500(2) | 10.4429(5) | 10.3657(3) |
| $x(O_1)$ | 0.328(3) | 0.333(1) | 0.334(1) | 0.334(1) | 0.340(9) |
| B(A) (Å$^2$) | 0.84(7) | 1.28(6) | 1.35(6) | 2.18(8) | 1.41(6) |
| B(B,B') (Å$^2$) | 0.17(9) | 0.11(8) | 0.60(8) | 1.23(10) | 0.14(7) |
| *Statistical parameters* | | | | | |
| N-P+C | 1431 | 1531 | 1800 | 2779 | 2742 |
| $R_{wp}$ | 16.5 | 13.4 | 12.5 | 15.8 | 8.36 |
| $R_{exp}$ | 9.71 | 9.65 | 8.88 | 14.44 | 7.9 |
| $\chi^2$ | 2.89 | 1.94 | 1.99 | 1.19 | 1.12 |
| $R_{Bragg}$ | 2.74 | 2.44 | 2.47 | 2.81 | 1.80 |
| *Interatomic distances (Å)* | | | | | |
| Ln-O1 (x6) | 2.606(7) | 2.583(6) | 2.536(6) | 2.529(6) | 2.473(5) |
| Ln-O2 (x2) | 2.286(1) | 2.281(1) | 2.262(1) | 2.261(1) | 2.244(1) |
| B-O1 (x6) | 2.039(5) | 2.044(5) | 2.044(5) | 2.046(5) | 2.055(4) |
| *Bonding angles (°)* | | | | | |
| O1-Ln-O1 | 116.7(2) | 116.5(2) | 116.1(2) | 116.1(2) | 115.4(4) |
| O1-A-O2(a) | 63.2(3) | 63.5(4) | 63.9(4) | 63.9(4) | 64.6(2) |
| O1-A-O2(b) | 79.5(1) | 79.1(1) | 78.5(1) | 78.4(1) | 77.4(1) |
| O1-A-O2(c) | 100.5(4) | 100.8(4) | 101.5(3) | 101.6(3) | 102.6(3) |

*† only known phase; previously reported value a = 10.534 (ref.4)*

```
 1
 2
 3
 4
 5
 6
 7
 8
 9
10
11
12
13
14
15
16
17
18
19
20
21
22
23
24
25
26
27
28
29
30
31
32
33
34
35
36
37
38
39
40
41
42
43
44
45
46
47
48
49
50
51
52
53
54
55
56
57
58
59
60
61
62
63
64
65
```